\documentclass{article}
\usepackage{spconfa4,amsmath,graphicx}
\usepackage{makecell}
\usepackage{multirow}
\usepackage{booktabs}
\usepackage{amsfonts}
\usepackage{cite}
\usepackage{bm}
\usepackage{svg}
\usepackage{gensymb}
\usepackage{circledsteps}
\usepackage[subtle,title=normal,sections=normal,margins=normal,mathdisplays=normal,mathspacing=normal, bibliography=normal]{savetrees}
\usepackage{hyperref}
\usepackage{blindtext}
\usepackage{xcolor}

\title{Flexible Multi-Channel Target Speaker Extraction Using \\ Geometry-Conditioned Spatially Selective Non-linear Filters}
\name{Jiatong Li, Wiebke Middelberg, Simon Doclo\thanks{This work was funded by the Deutsche Forschungsgemeinschaft (DFG, German Research Foundation) under Germany's Excellence Strategy - EXC 2177/2 - Project ID 390895286, Project ID 352015383 - SFB 1330 B2, and Project ID 568930428. Simulations were conducted on the HPC cluster ROSA funded by the DFG under INST 184/225-1 FUGG.}}
\address{Dept. of Medical Physics and Acoustics and Cluster of Excellence Hearing4all,\\ Carl von Ossietzky Universität Oldenburg, Oldenburg, Germany\\
\{\href{mailto:jiatong.li@uni-oldenburg.de}{jiatong.li},\href{mailto:wiebke.middelberg@uni-oldenburg.de}{wiebke.middelberg}, \href{mailto:simon.doclo@uni-oldenburg.de}{simon.doclo}\}@uni-oldenburg.de}
\begin{document}
\ninept
\maketitle
\begin{abstract}
Recently, a spatially selective non-linear filter (SSF) has been proposed for target speaker extraction, using the target direction-of-arrival (DOA) as a spatial cue. Since learned intermediate features are tied to the microphone geometry, the performance of the SSF degrades significantly when evaluated on mismatched array geometries. In this paper, we propose a geometry-conditioned SSF (GC-SSF), which incorporates a geometry-conditioning branch based on FiLM layers. Furthermore, we propose a feature that jointly encodes the DOA and the microphone positions (DOA-MPE). The conditioning branch modulates the intermediate feature maps of the SSF using the DOA-MPE feature to capture the spatial relationship between the microphone positions and the target speaker. Experimental results across circular, uniform linear, and random microphone arrays show that the proposed GC-SSF generalizes better to mismatched geometries while maintaining high spatial selectivity, demonstrating its ability to effectively adapt the filtering process to different array geometries.
\end{abstract}
\begin{keywords}
spatially selective non-linear filter, geometry conditioning, microphone array processing, target speaker extraction
\end{keywords}
\section{Introduction}
\label{sec:intro}
Extracting a target speaker from a mixture of speakers and background noise remains a fundamental challenge in acoustic signal processing\cite{overview}. To discriminate the target speaker from the interfering speakers, various cues have been proposed, such as enrollment utterances \cite{8462661,sinha2024variants}, visual information \cite{9380418, 10094306}, and spatial features \cite{8540037, beamformer_guided, tesch2024ssf, 11230948, shetu2025ganbasedmultimicrophonespatialtarget,10888158}. In this paper, we focus on the spatially selective non-linear filter (SSF) proposed in \cite{tesch2024ssf}. The SSF utilizes the target direction-of-arrival (DOA) as a spatial cue and employs Long Short-Term Memory (LSTM) layers to estimate a complex-valued mask that is applied to the reference microphone signal to extract the target speaker.
%, which operates on multichannel speech signals and utilizes the target direction-of-arrival (DOA) as a spatial cue to steer a non-linear filter.

Despite its effectiveness, a limitation of the SSF is its inherent dependency on a fixed microphone array geometry. Since the SSF is typically trained on a specific microphone array, the learned intermediate features are tied to its specific array geometry. Consequently, the performance degrades significantly when the SSF system is evaluated on mismatched array geometries. Although a meta-learning approach \cite{mannanova2024meta} attempts to address this limitation via fine-tuning, it often lacks scalability for diverse real-world geometries. Alternatively, various geometry-agnostic systems \cite{luo2020TAC,jukic2023TAC,Tammen2024TAC,lee24g_interspeech, AmbiDrop2026, eigenbeam2026,flexio2026} have been proposed, which are not tied to a specific array geometry. However, these systems are designed for noise reduction or blind speaker separation rather than DOA-based target speaker extraction. Although geometry-conditioning systems have been proposed for DOA estimation \cite{kowalk2023geometry,pe} and blind speaker separation \cite{fan2024geoemtry}, to the best of our knowledge geometry-conditioning has not been considered for DOA-based target speaker extraction.
\begin{figure}[t]
%\frac{\langle \hat{\mathbf{s}}, \mathbf{s} \rangle \mathbf{s}}{\|\mathbf{s}\|^2}

  \centering
    \includegraphics[width= 0.95\linewidth]{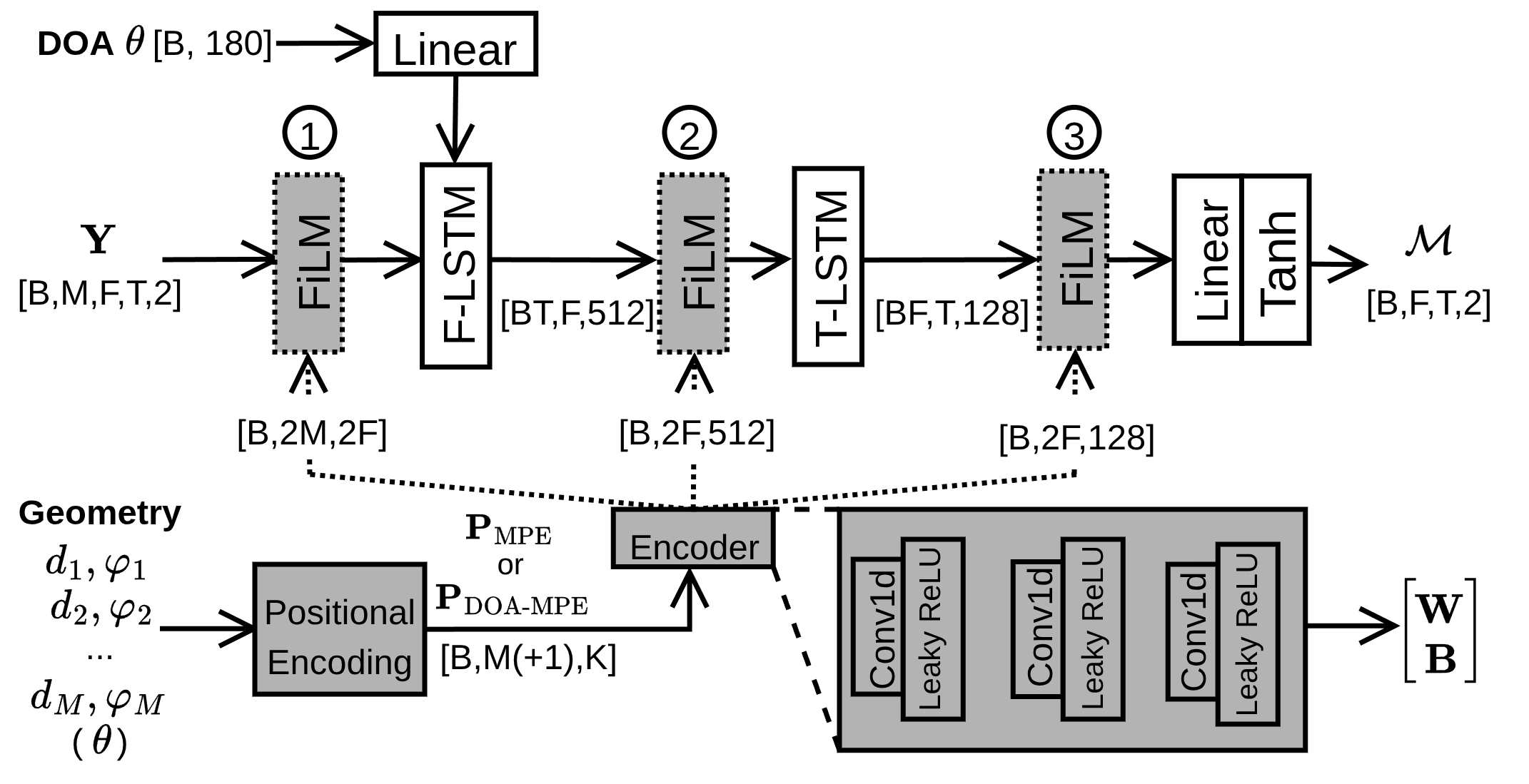}
%  \vspace{1.5cm}
  \label{fig:system}
%
% \vspace{-2ex}
 \caption{Overview of the proposed geometry-conditioned SSF. White blocks represent the baseline SSF architecture, while gray blocks denote the proposed geometry-conditioning branch. The output of the encoder is integrated into the SSF system via a FiLM layer at one of three points of injection (POI \Circled{1}, \Circled{2}, or \Circled{3}).}
\label{fig:diagram}
\end{figure}
% {\color{red} I am not sure where to put this, but we definitely have to mention the geometry-agnostic (TAC) processing \cite{casebeer2018multiview,luo2020TAC,jukic2023TAC,Tammen2024TAC} and more geometry-informed literature like \cite{kowalk2023geometry,fan2024geoemtry}. Of course we must also discuss the downsides (like Luo is speaker separation, i.e., needs subsequent cue for target speaker selection. Or Jukic/Tammen is single speaker in noise.}

In this paper, we propose a geometry-conditioned spatially selective filter (GC-SSF), which uses the microphone array geometry as a conditioning input to achieve robustness across different array geometries, assuming a fixed number of microphones. 
The main contributions are: 1) integration of a geometry-conditioning branch into the baseline SSF (see Fig.~\ref{fig:diagram}), using a Feature-wise Linear Modulation (FiLM) layer \cite{film} to modulate intermediate feature maps from the SSF system, 2) a DOA-Microphone Positional Encoding (DOA-MPE) feature, which effectively represents the spatial relationship between the microphone positions and the target speaker to provide the important geometric context for DOA-steered target speaker extraction.
In the experimental evaluation, we compare the proposed GC-SSF system, trained on random arrays, against baseline SSF systems, trained either on a fixed circular array or on random arrays. The performance is evaluated on both matched and mismatched array geometries. Experimental results show that the proposed GC-SSF consistently outperforms the baseline SSF trained on random arrays across all evaluated geometries. While the baseline SSF trained on the circular array shows the best performance for the matched circular array geometry, a substantial performance degradation is observed for mismatched geometries. In contrast, the GC-SSF effectively bridges this performance gap, achieving robust performance across all considered geometries.
%demonstrating its improved generalization ability compared to the baseline SSF system.
Furthermore, a sensitivity analysis with respect to target DOA errors demonstrates that the proposed GC-SSF achieves both strong generalization across different array geometries and high spatial selectivity.
\section{Spatially Selective Non-linear Filter}
\label{sec:ssf}
In this section, we review the spatially selective non-linear filter (SSF) for target speaker extraction\cite{tesch2024ssf}, which serves as the baseline system.

In the short-time Fourier transform (STFT) domain, the observed noisy speech signal at the $m$-th microphone for frequency bin $f \in [1, F]$ and time frame $t \in [1, T]$ is denoted by $Y_m(f,t)$, where $F$ and $T$ denote the number of frequency bins and time frames, respectively. By stacking the signals from all $M$ microphones, the multichannel observed noisy speech vector $\mathbf{Y}(f,t) = [Y_1(f,t), Y_2(f,t), ... , Y_M(f,t)]^\text{T} \in \mathbb{C}^M$ is given by
\begin{equation}
\mathbf{Y}(f,t) = \mathbf{X}(f,t) + \mathbf{V}(f,t),
\end{equation}
where the target speech vector $\mathbf{X}(f,t) \in \mathbb{C}^M$ and the interfering speech vector $\mathbf{V}(f,t) \in \mathbb{C}^M$ are defined similar as $\mathbf{Y}(f,t)$.

The goal is to recover the reverberant target speech component $X_1(f,t)$ from the noisy speech vector $\mathbf{Y}(f,t)$, with the first microphone chosen as the reference microphone. To achieve this, the SSF estimates a single-channel complex mask $\mathbf{\mathcal{M}}(f,t)$, which is applied to the noisy speech at the reference microphone $Y_1(f,t)$, i.e.,
\begin{equation}
\hat{X}_1(f,t) = \mathcal{M}(f,t) Y_1(f,t).
\end{equation}

 The SSF estimates the mask $\mathcal{M}(f,t)$ from the noisy speech vector $\mathbf{Y}(f,t)$, conditioned on the target DOA $\theta$ (see Fig.~\ref{fig:diagram}). The SSF architecture comprises two Long Short-Term Memory (LSTM) layers to jointly process spatial, spectral, and temporal information. The first LSTM layer, the frequency-domain LSTM (F-LSTM), analyzes spatial and spectral information by encoding the noisy speech vector $\mathbf{Y}(f,t)$ into a high-dimensional feature vector. This is followed by a time-domain LSTM (T-LSTM), which models spatial and temporal dependencies by processing the F-LSTM outputs along the time dimension. Finally, a linear layer outputs the complex mask $\mathbf{\mathcal{M}}(f,t)$. 
To steer the SSF, the target DOA $\theta$ is mapped to a 180-dimensional one-hot vector with a $2^\circ$ resolution. This vector is projected via a linear layer to initialize the cell state of the F-LSTM, thereby conditioning the SSF on the target DOA. 

Typically, the SSF system is trained on a specific microphone array. This, however, makes the learned intermediate features inherently dependent on the microphone array considered during training. As a result, 
%even with a fixed number of microphones,
the system exhibits substantial performance degradation when applied to mismatched array geometries.

\section{PROPOSED geometry-conditioned \\ Spatially selective Non-linear Filter}
\label{sec: gc-ssf}
To improve the generalization ability of the SSF system across different microphone array geometries for a fixed number of microphones, we propose to incorporate a geometry-conditioning branch into the SSF (see Fig. \ref{fig:diagram}). This branch first transforms the microphone array geometry and the target DOA to spatial features using a positional encoding scheme (see Section \ref{sec:pe}). These spatial features are then processed by an encoder, whose output is used to modulate the intermediate feature maps of the SSF via a FiLM layer (see Section \ref{sec:film}).
\subsection{Positional Encoding}
\label{sec:pe}
To represent the microphone array geometry for DNN processing, we utilize the Microphone Positional Encoding (MPE) scheme \cite{pe}. The MPE encodes the position of the $m$-th microphone using their polar coordinates $(\varphi_m, d_m)$ relative to the array centroid into high-dimensional sinusoidal features without requiring learnable parameters, i.e.,
\begin{equation}
\mathbf{p}_m = \alpha\,d_m\, \begin{bmatrix}\cos (2\pi \sigma \mathbf {v}+\varphi_{m}) \\ \sin (2\pi \sigma \mathbf {v}+\varphi _{m})\end{bmatrix}\,\in \mathbb {R}^{K},
\end{equation}
where $\mathbf {v} = \frac{2}{K}[0, 1, ..., \frac{K}{2}-1]^\text{T} \in  \mathbb {R}^{\frac{K}{2}}$ is a constant vector, and $\alpha, \sigma$, and $K$ are constant hyperparameters. Here, $d_m$ represents the Euclidean distance from the $m$-th microphone to the array centroid. The azimuth angles $\varphi_m$ are defined counter-clockwise relative to the reference axis passing through the array centroid and the reference microphone (see Fig. \ref{fig:micgeo}). As described in \cite{pe}, by employing these sinusoidal features, the MPE explicitly reflects the phase-related relationships between microphones that are critical for spatial filtering. This is a property not easily captured by simply concatenating microphone coordinates \cite{kowalk2023geometry}. 
%Furthermore, the transformation of microphone coordinates into a high-dimensional feature enables the utilization of more advanced DNN architectures than linear layers.
By stacking the positional encoding features $\mathbf{p}_m$ from all $M$ microphones, the positional encoding feature for the whole microphone array $\mathbf{P}_{\text{MPE}}$ is defined as,
\begin{equation}
\mathbf{P}_{\text{MPE}} = [\mathbf{p}_1, \mathbf{p}_2, ... \mathbf{p}_M]\, \in \mathbb {R}^{K\times M}.
\end{equation}

While $\mathbf{P}_{\text{MPE}}$ characterizes the array geometry, it does not explicitly capture the spatial relationship between the microphone positions and the target source. To facilitate the network in learning the relationship between the array geometry and the target DOA $\theta$, which is anticipated to be important for DOA-steered target speaker extraction, we propose an augmented feature, DOA-MPE:
\begin{align*}
    \mathbf{P}_{\text{DOA-MPE}} &= [\mathbf{P}_{\text{MPE}}, \mathbf{p}_{\text{DOA}}]\, \in \mathbb {R}^{K\times (M+1)}, \tag{5}\\
  \mathbf{p}_{\text{DOA}} = \,&\alpha\, \begin{bmatrix}\cos (2\pi \sigma \mathbf {v}+\theta) \\ \sin (2\pi \sigma \mathbf {v}+\theta)\end{bmatrix}\,\in \mathbb {R}^{K}, \tag{6}
\end{align*}
where the target DOA $\theta$ is defined relative to the same reference axis as the microphone azimuth angle $\varphi_m$. It should be noted that since the distance between the array centroid and the target speaker is typically unknown in practice, we only incorporate the target DOA $\theta$ into $\mathbf{p}_{\text{DOA}}$.
% \begin{equation}
% \mathbf{p}_{\text{DOA}} = \alpha \begin{bmatrix}\cos (2\pi \sigma \mathbf {v}+\theta) \\ \sin (2\pi \sigma \mathbf {v}+\theta)\end{bmatrix} \in \mathbb {R}^{K},
% \end{equation}
% \begin{equation}
% \mathbf{P}_{\text{DOA-MPE}} = [\mathbf{P}_{\text{MPE}}, \mathbf{p}_{\text{DOA}}] \in \mathbb {R}^{K\times (M+1)}.
% \end{equation}
\subsection{Conditioning using FiLM}
\label{sec:film}
To condition the target speaker extraction on the microphone array geometry, we utilize a FiLM layer \cite{film} to modulate the intermediate feature maps $\mathbf{O}(t)$ of the SSF at time frame $t$, i.e.,
\begin{equation}
    \text{FiLM}(\mathbf{O}(t))=\mathbf{W}\odot\mathbf{O}(t)+\mathbf{B},
    \tag{7}
\end{equation}
\begin{equation}
    \begin{bmatrix}\mathbf{W}\\\mathbf{B}\end{bmatrix} = \text{Encoder}(\mathbf{P}), \tag{8}
\end{equation}
where $\odot$ denotes element-wise multiplication.
The conditioning parameters, the scaling matrix $\mathbf{W}$ and the bias matrix $\mathbf{B}$, are estimated by an encoder from the input positional encoding feature $\mathbf{P} \in \{ \mathbf{P}_{\text{MPE}}, \mathbf{P}_{\text{DOA-MPE}} \}$.
In the proposed system, the parameters $\mathbf{W}$ and $\mathbf{B}$ are time-invariant, as the array geometry and the target DOA are assumed to be static. 
For the encoder, we employ one-dimensional convolutional (Conv1d) layers followed by Leaky ReLU activation functions. The Conv1d layers are suitable for this task as they enable the network to effectively exploit phase-related relationships across microphone and DOA dimensions in the positional encoding feature $\mathbf{P}_\text{DOA-MPE}$. As illustrated in Fig.~\ref{fig:diagram}, we investigate three potential points of injection (POIs) for the conditioning mechanism, denoted as \Circled{1}, \Circled{2}, and \Circled{3}. Depending on the selected POI, the output dimensionality of the encoder is adjusted to match the dimensions of the corresponding intermediate feature maps $\mathbf{O}(t)$. This flexibility allows us to evaluate the effect of applying geometric conditioning at different stages of the spatial filtering process.
\section{Experiments}
This section first presents the experimental setup, including the training and evaluation datasets, the network structure, and the training procedure.
Then, the experimental results are presented and discussed, evaluating the performance, generalization ability, and the spatial selectivity of the proposed GC-SSF system compared with the baseline SSF systems.
\subsection{Datasets}
\begin{figure}[t]
    \centering
    % Left Side: The Image
    \begin{minipage}[b]{0.58\linewidth}
        % \centering
        \hspace{-7mm}
        \includegraphics[width=5.0cm]{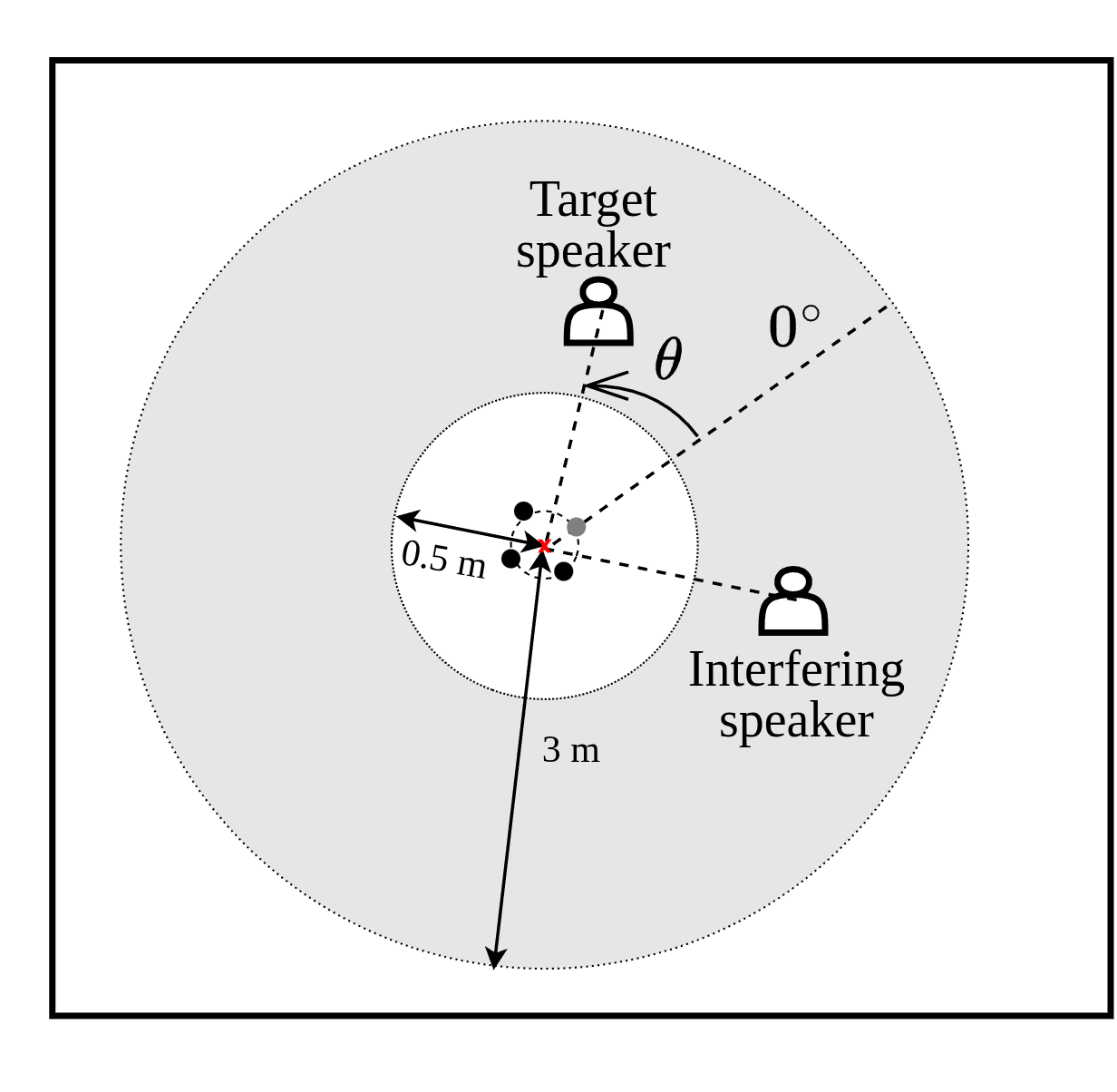}
        % \vspace{2pt} % Adds a little breathing room
        % \\ (a) Acoustic setup
    \end{minipage}
    % \hfill
    % Right Side: The Table
    \begin{minipage}[b]{0.30\linewidth}
        % \centering
        \hspace{-6mm}
        \small % Slightly smaller text to ensure it fits the 0.30 width
        \begin{tabular}{ll}
            \toprule
            \multicolumn{2}{c}{\textbf{Room characteristics}} \\ 
            \midrule
            Width  & 2.5--5 m \\
            Length & 3--9 m   \\
            Height & 2.2--3.5 m \\
            $T_{60}$    & 0.2--0.5 s \\
            \bottomrule
        \end{tabular}
        \vspace{1cm}
        % \\ (b) Parameters
    \end{minipage}
    \vspace{-3mm}
    \caption{Illustration of the simulation setup and the room characteristics. The target speaker and the interfering speaker are located in the gray region.}
    \label{fig:setups}
\end{figure}

\begin{figure}[t]
%\frac{\langle \hat{\mathbf{s}}, \mathbf{s} \rangle \mathbf{s}}{\|\mathbf{s}\|^2}
\begin{minipage}[b]{0.30\linewidth}
  % \centering
  \centerline{\includegraphics[width=2.7cm]{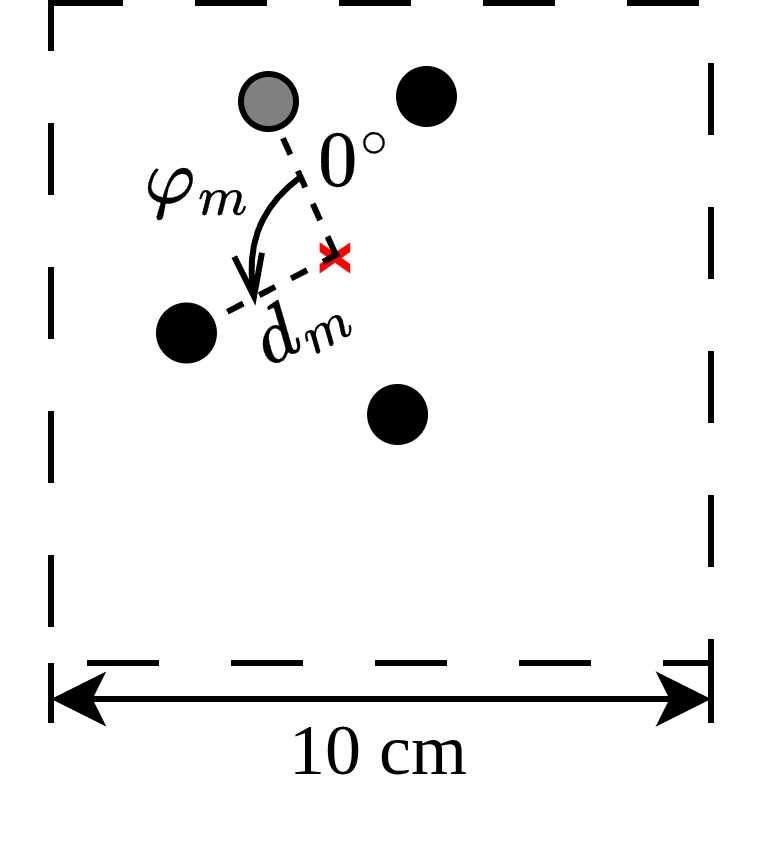}}
 \vspace{-0.3cm}
  \centerline{(a) Random}
  \label{figa}
\end{minipage}
\hfill
\begin{minipage}[b]{0.30\linewidth}
  % \centering
  \centerline{\includegraphics[width=2.5cm]{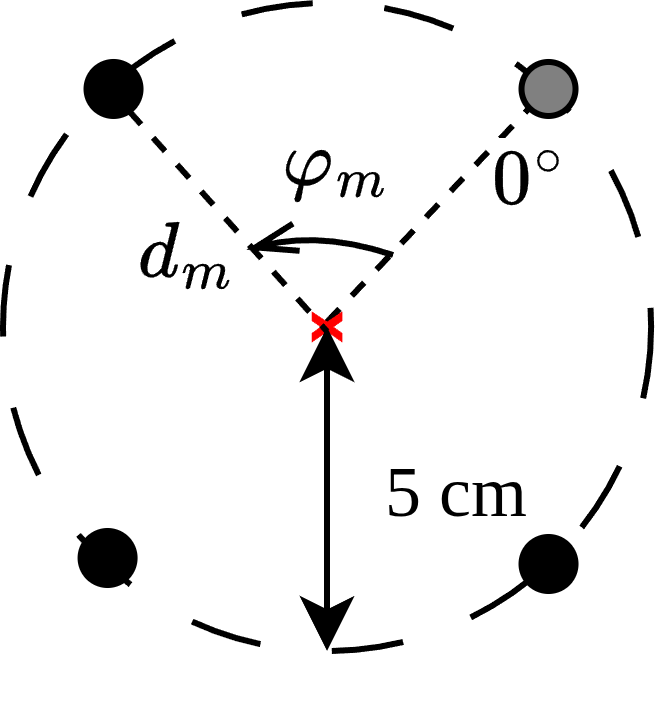}}
 % \vspace{0.2cm}
  \centerline{(b) Circ}
    \label{figb}
\end{minipage}
\hfill
\begin{minipage}[b]{0.30\linewidth}
  % \centering
\centerline{\hspace{0.5cm}\includegraphics[width=2.3cm]{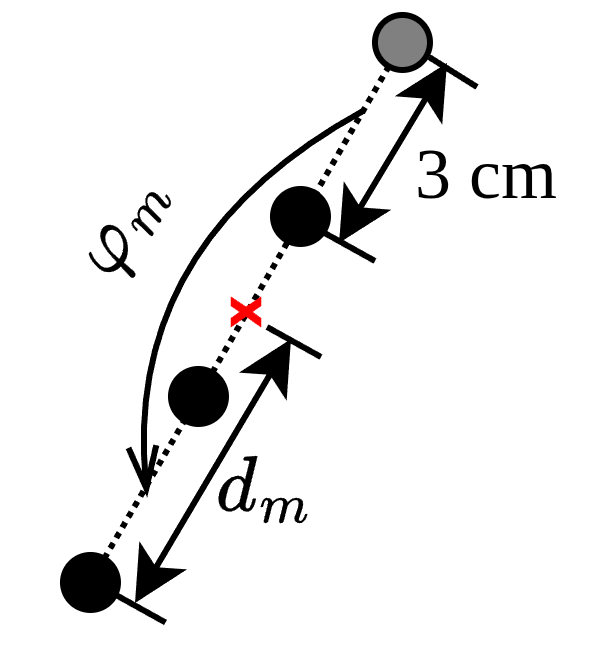}}
 % \vspace{0.3cm}
  \centerline{(c) ULA}
    \label{figc}
\end{minipage}
% \vspace{-2mm}
 \caption{Considered random, circular and ULA array geometries with $M=4$ microphones. Gray circle denotes the reference microphone. Red cross denotes the array centroid.}
\label{fig:micgeo}
\end{figure}

To evaluate the performance of the proposed GC-SSF system, acoustic environments were simulated using the Pyroomacoustics library \cite{Scheibler2018pyroomacoustics}. The room dimensions and the reverberation time $T_{60}$ were uniformly sampled from the ranges specified in Fig.~\ref{fig:setups} (right).
In order to investigate the generalization ability of the system to mismatched array geometries, three array geometries with $M=4$ microphones were considered: Circular (Circ), Uniform Linear (ULA), and Random arrays (see Fig.~\ref{fig:micgeo}). These arrays were randomly placed in the horizontal plane at a height of $1.6$~m with a random rotation. For each array geometry, the first microphone was chosen as the reference microphone. For the random array, the microphone positions were randomly sampled within a square region of $10 \times 10$~cm$^2$.
As illustrated in Fig.~\ref{fig:setups} (left), one target speaker and one interfering speaker were placed at a height of $1.6$~m. The angular separation between two speakers was constrained to a minimum of 20\degree. The acoustic datasets were generated using speech signals from the Wall Street Journal (WSJ0) corpus \cite{wsj0data}, sampled at $16$~kHz.  Following the data split defined in \cite{tesch2024ssf}, the corpus is split into training, validation, and evaluation sets, ensuring disjoint sets of speakers. The training and validation sets comprise 30\,h and 1\,h of data, respectively, with a utterance duration of 3\,s. Similarly, 1\,h of data was generated for each array geometry for evaluation. The mixture signals were generated such that the ratio between the power of the reverberant target speech and the power of the reverberant interfering speech at the reference microphone was between -5\,dB and 10\,dB.
\subsection{Network and Training}
Following the framework in \cite{tesch2024ssf}, the STFT is computed using a frame length of 512 with 50\% overlap and a square-root Hann window for both analysis and synthesis. The baseline SSF model, illustrated in Fig.~\ref{fig:diagram}, adopts the network architecture described in \cite{tesch2024ssf}. For the positional encoding block, the hyperparameters were set to: $\alpha=7, \sigma=4$, and $K=514$.
The encoder consists of three Conv1d layers with a kernel size of 5, each followed by a LeakyReLU activation function. The first two layers contain 64 and 128 output channels, respectively, while the third Conv1d layer contains $2M$ output channels for POI \Circled{1} and $2F$ output channels for POIs \Circled{2} and \Circled{3}.
To ensure the positional encoding feature ($\mathbf{P}_{\text{MPE}}$ or $ \mathbf{P}_{\text{DOA-MPE}}$) matches the dimensions of the intermediate feature map at the selected POI, the strides for the Conv1d layers are set to $(1, 1, 1)$ for POIs \Circled{1} and \Circled{2}, and $(2, 2, 1)$ for POI \Circled{3}. The encoder output dimensions for each POI are illustrated in Fig.~\ref{fig:diagram}.
% \begin{table}
%     \centering
%     \begin{tabular}{c|c|c|c}
%         POI & Kernel size & Output channel & Strides \\
%         \hline
%         \Circled{1} & 5 & (64, 128, 2M) & (1,1,1) \\
%         \Circled{2} & 5 & (64, 128, 2F) & (1,1,1) \\
%         \Circled{3} & 5 & (64, 128, 2F) & (2,2,1) \\
%     \end{tabular}
%     \caption{Dummy table with overview - not going to be in paper.}
%     \label{tab:my_label}
% \end{table}

The training procedure follows the strategy in \cite{tesch2024ssf}. All the networks were trained to minimize a loss function, which combines time-domain and frequency-domain loss terms, i.e.,
\begin{equation}
\mathcal{L}\hspace{-0.5mm}=\hspace{-0.5mm}\frac{\beta}{N} \hspace{-1mm}\sum_{n=1}^{N}\hspace{-0.5mm}|x_1(n)\hspace{-0.2mm} -\hspace{-0.2mm} \hat{x}_1(n)| \hspace{-0.1mm} + \hspace{-0.1mm}\frac{1}{TF}\hspace{-1mm}\sum_{t=1}^{T}\hspace{-0.7mm}\sum_{f=1}^{F}\hspace{-0.5mm}\left| |X_1(f,t)| \hspace{-0.8mm}- \hspace{-0.8mm}|\hat{X}_1(f,t)| \right|, \hspace{-1mm}\tag{9}\label{loss_func}\end{equation}
% for the freq loss part, no resynthesis
where $x_1(n)$ denotes the reverberant time-domain target speech signal at the reference microphone, and $\hat{x}_1(n)$ represents its estimated time-domain signal reconstructed via inverse STFT and overlap-add. Here, $n$ denotes the time sample index and $N$ denotes the number of samples. The scalar weighting factor $\beta$ is set to $\beta = 10$ following the configuration in \cite{tesch2024ssf}. All networks were trained for 500 epochs with a batch size of 16. Optimization was performed using the Adam optimizer with an initial learning rate of 0.001. A multi-step learning rate scheduler was employed, decaying the learning rate by a factor of 0.75 every $50$ epochs, up to epoch $400$. To ensure stable convergence, the $l_2$-norm of gradients are clipped to 1.
\subsection{Experimental Results}
\begin{figure}[t]

\centering
\includegraphics[width=0.95\linewidth]{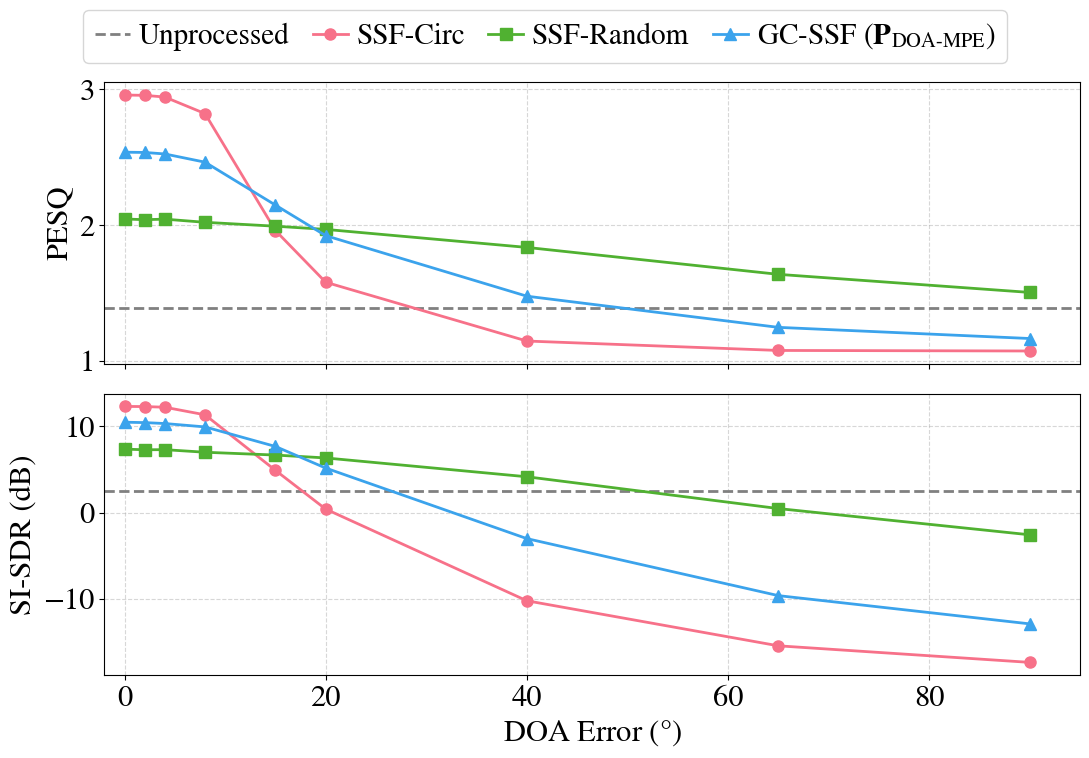}

\caption{Average PESQ and SI-SDR (dB) of the baseline SSF (trained on circular or random arrays) and proposed GC-SSF (trained on Random arrays) with positional encoding feature $\mathbf{P}_{\text{DOA-MPE}}$ at POI \Circled{2} for different target DOA errors. Systems are evaluated on the circular array.} 
\label{fig:DOA_curve}
\end{figure}

In this section, we provide an experimental evaluation of the GC-SSF. First, we investigate the optimal configuration by comparing different positional encoding features ($\mathbf{P}_{\text{MPE}}$ and $\mathbf{P}_{\text{DOA-MPE}}$) and POIs (see Table \ref{tab: poi}). Then, we compare the performance and generalization ability of the GC-SSF with baseline SSF systems (see Table \ref{tab: comparebaseline}). In the end, we analyze the spatial selectivity of the systems by evaluating their sensitivity to target DOA errors (see Fig. \ref{fig:DOA_curve}). The performance is evaluated using the wide-band Perceptual Evaluation of Speech Quality (PESQ) \cite{Rix2001PESQ} and the Scale Invariant Signal-to-Distortion Ratio (SI-SDR) \cite{sisdr}, with the reverberant target speech at the reference microphone as the clean reference signal.

Table \ref{tab: poi} shows the average PESQ results for the GC-SSF system configured with different positional encoding features across the three considered POIs. In this experiment, the models are trained on Random arrays and evaluated on both Circ and Random geometries. As shown in the table, the proposed feature $\mathbf{P}_{\text{DOA-MPE}}$ consistently outperforms the feature $\mathbf{P}_{\text{MPE}}$ for all considered POIs and evaluation geometries. This suggests that the joint encoding of the microphone positions and the target DOA enables the conditioning branch to effectively model the spatial relationship between the microphones and the target source. Regarding the choice of POI, integrating the conditioning at the output of the T-LSTM (POI \Circled{3}) yields lower performance compared to earlier points of injections (POI \Circled{1} and \Circled{2}), suggesting that %early-stage conditioning ensures geometric constraints are deeply integrated throughout the speech extraction process, whereas 
late-stage injection limits the network's ability to adapt its internal representations to different array geometries. Since the feature $\mathbf{P}_{\text{DOA-MPE}}$ combined with POI \Circled{2} achieves the highest PESQ score, it is utilized as the default configuration for the GC-SSF in the remaining experiments.

Table \ref{tab: comparebaseline} compares the proposed GC-SSF trained on Random arrays with the baseline SSF systems trained on either Circ array (SSF-Circ) or Random array (SSF-Random). The baseline SSF-Circ shows the best performance in the matched Circ dataset but fails to generalize to mismatched ULA and Random geometries, resulting in PESQ scores lower than those of the unprocessed mixtures. This indicates that the system learns spatial features tied specifically to the Circ geometry. In contrast, the baseline SSF-Random system shows more consistent performance across different geometries but suffers from overall performance degradation. This can be explained by the inconsistent mapping between the input DOA and the array-dependent phase-related relationships in observed multichannel signals, which is unresolvable in random arrays without explicit geometry information. Compared to the SSF-Circ and the SSF-Random, the proposed GC-SSF addresses these limitations. The GC-SSF surpasses the SSF-Random by approximately 0.45 in PESQ across all evaluated geometries and achieves a substantial improvement of up to 1.25 in PESQ over the SSF-Circ in mismatched array geometries. While the baseline SSF-Circ achieves better performance than our proposed GC-SSF in the matched Circ geometry, these results demonstrate that the proposed GC-SSF system achieves better generalization ability across different array geometries.
\begin{table}[t]	
\setlength{\tabcolsep}{5pt}
% \centering
\caption{{Average PESQ of GC-SSF with different positional encoding features for three POIs (\Circled{1}, \Circled{2} and \Circled{3}), Systems are trained with random arrays and evaluated on circular and random arrays.}}
\label{tab: poi}
% \centering
% \small
\vspace{1mm}
\begin{tabular}{c|cc|cc|cc}
	\toprule
	% Add the first column for vertical text
	 \multirow{2}*{Feature} &\multicolumn{2}{c|}{POI \Circled{1}}& \multicolumn{2}{c|}{POI \Circled{2}} & \multicolumn{2}{c}{POI \Circled{3}}\\
	 & Circ & Random & Circ& Random& Circ& Random \\
	%\hline
    \cline{1-7}
    % MPos &\makecell{1.72 \\ ($\pm\,$0.59)} & \makecell{1.70 \\ ($\pm\,$0.58)}  & \makecell{2.09 \\ ($\pm\,$0.65)} & \makecell{2.02 \\ ($\pm\,$0.67)} & \makecell{1.99 \\ ($\pm\,$0.62)} & \makecell{1.89 \\ ($\pm\,$0.63)} \\
    % DOA-MPos &\makecell{2.51 \\ ($\pm\,$0.66)} & \makecell{2.43 \\ ($\pm\,$0.64)} & \makecell{2.53 \\ ($\pm\,$0.63)} & \makecell{2.46 \\ ($\pm\,$0.63)} & \makecell{2.12 \\ ($\pm\,$0.64)} & \makecell{2.15 \\ ($\pm\,$0.65)}  \\
    % % FD-DOA-MPos & \makecell{2.71 \\ ($\pm\,$0.60)} & \makecell{2.65 \\ ($\pm\,$0.62)} & \makecell{2.33 \\ ($\pm\,$0.64)} & \makecell{2.30 \\ ($\pm\,$0.63)} \\
    % % \cline{1-10}
    $\mathbf{P}_{\text{MPE}}$ &1.72 & 1.70  & 2.09 & 2.02  & 1.99  & 1.89  \\
    $\mathbf{P}_{\text{DOA-MPE}}$ &2.51  & 2.43  & \textbf{2.53}  & \textbf{2.46} & 2.12  & 2.15 \\

\bottomrule                             
\end{tabular}	
\end{table}
\begin{table}[t]	
\vspace{-2ex}
\setlength{\tabcolsep}{3pt}
\centering
\caption{{Average PESQ (with standard deviation) of the baseline SSF (trained on circular or random arrays) and the proposed GC-SSF (trained on random arrays) with positional encoding feature $\mathbf{P}_{\text{DOA-MPE}}$ at POI \Circled{2}. Systems are evaluated on circular, ULA, and random arrays.}}
\label{tab: comparebaseline}
\centering
\small
\vspace{1mm}
\begin{tabular}{c|c|ccc}
	\toprule
	% Add the first column for vertical text
	 \multirow{2}*{System} & \multirow{2}*{Train} &\multicolumn{3}{c}{Evaluation} \\
	 & & Circ & ULA & Random \\
	%\hline
    \cline{1-5}
     Unprocessed & - & \makecell{1.38 \\ ($\pm\,$0.28)} & \makecell{1.39 \\ ($\pm\,$0.29)} & \makecell{1.36 \\ ($\pm\,$0.27)} \\
    \hline
    SSF-Circ & Circ  & \makecell{\textbf{2.95} \\ ($\pm\,$0.59)} & \makecell{1.16 \\ ($\pm\,$0.28)} & \makecell{1.20 \\ ($\pm\,$0.35)} \\
    SSF-Random & Random &\makecell{2.04 \\ ($\pm\,$0.67)} & \makecell{2.02 \\ ($\pm\,$0.66)} & \makecell{1.93 \\ ($\pm\,$0.67)} \\
	\hline
    % (3) GC-SSF + $\mathbf{P}_{\text{MPE}}$ & Random & \makecell{2.09 \\ ($\pm\,$0.65)} & \makecell{2.12 \\ ($\pm\,$0.68)} & \makecell{2.02 \\ ($\pm\,$0.67)} \\
    GC-SSF ($\mathbf{P}_{\text{DOA-MPE}}$) & Random & \makecell{2.53 \\ ($\pm\,$0.63)} & \makecell{\textbf{2.41} \\ ($\pm\,$ 0.66)} & \makecell{\textbf{2.46} \\ ($\pm\,$0.63)} \\

    % & (5) FD-DOA-MPos  & \makecell{2.71 \\ ($\pm\,$0.60)} & \makecell{2.45 \\ ($\pm\,$0.69)} & \makecell{2.65 \\ ($\pm\,$0.62)} \\

    % \cline{1-10}

\bottomrule                             
\end{tabular}	
\end{table}

To investigate the spatial selectivity of the systems, we evaluate them on the Circ array under different target DOA errors. The results are illustrated in Fig.~\ref{fig:DOA_curve}. In general, the PESQ and SI-SDR scores of all systems decrease as the target DOA error increases. Specifically, for the baseline SSF-Circ, the PESQ and SI-SDR scores decrease significantly when the target DOA error exceeds 15\degree, which indicates that the baseline SSF-Circ effectively utilizes the input DOA information and achieves high spatial selectivity. The underlying reason is that the fixed geometry in training allows the network to learn a direct mapping between the inter-channel phase relationships and the provided DOA, resulting in a trade-off between high spatial selectivity and robustness to target DOA errors. In contrast, the baseline SSF-Random exhibits considerably lower sensitivity to target DOA errors than the SSF-Circ in terms of both PESQ and SI-SDR. This implies that, without a fixed geometric context or explicit conditioning, the system cannot learn the relationship between the array geometry and the input DOA. Consequently, the system fails to properly use the given DOA information, resulting in low spatial selectivity and degraded performance. Compared with these baseline systems, the proposed GC-SSF achieves higher SI-SDR and PESQ scores than the SSF-Random for DOA errors below 20\degree, while showing a comparable spatial selectivity to the SSF-Circ. This shows that the GC-SSF effectively uses the input DOA information, despite being trained on Random arrays. Overall, these results demonstrate that explicit geometry conditioning enables the GC-SSF to achieve both high spatial selectivity and generalization ability across different array geometries.
\section{Conclusions}
\label{sec:conclusions}
In this paper, we proposed the GC-SSF system, designed to achieve robust target speaker extraction across different array geometries for a fixed number of microphones. The proposed system extends the baseline SSF by incorporating an explicit geometry-conditioning branch via a FiLM layer and a proposed DOA-MPE feature to represent the spatial relationship between the microphone positions and the target source.
Experimental results demonstrate that while a baseline SSF trained on a fixed geometry achieves competitive performance in matched conditions, the proposed GC-SSF trained on random arrays shows better generalization ability across mismatched array geometries. A sensitivity analysis to target DOA errors further shows that the proposed GC-SSF maintains high spatial selectivity without sacrificing generalization ability across different array geometries.
While the current architecture is designed for a fixed number of microphones, future work will investigate architectures that are independent of the number of microphones, thereby enabling the deployment of the GC-SSF in ad-hoc acoustic sensor networks.

\bibliographystyle{IEEEbib}
\bibliography{refs}

\end{document}